\documentclass[11pt]{article}

\usepackage{graphicx}
\usepackage{float}
\usepackage{booktabs}
\usepackage{amsmath,amsfonts,bbm,bm,mathtools,relsize,exscale}
\usepackage{dsfont}
\usepackage[OT1]{fontenc}
\usepackage[utf8]{inputenc}
\usepackage[inline,shortlabels]{enumitem}

\makeatletter
\@ifclassloaded{biom}{
\bibliographystyle{biom} 
}{
\bibliographystyle{biometrika}
}
\makeatother

\graphicspath{ {figs/} }

\usepackage{refcount,nameref,zref-xr,zref-user} 
\zxrsetup{toltxlabel} 
\usepackage{url}

\let\href\undefined
\usepackage[colorlinks,citecolor=blue,urlcolor=blue]{hyperref}

\newcommand{\titlepaper}{
On the use of cross-fitting in causal machine learning with correlated units
}
\newcommand{\abstracttext}{
  In causal machine learning, the fitting and evaluation of nuisance models are often performed on separate partitions, or folds, of the observed data. This technique, called cross-fitting, eliminates bias introduced by the use of black-box predictive algorithms. When study units may be correlated, such as in spatial, clustered, or time-series data, investigators often design bespoke forms of cross-fitting to minimize correlation between folds. We prove that, perhaps contrary to popular belief, this is typically unnecessary: performing cross-fitting as if study units were independent still eliminates key bias terms even when units may be correlated. In simulation experiments with various correlation structures, we show that causal machine learning estimators achieve the same or improved bias and precision under cross-fitting that ignores correlation compared to techniques striving to eliminate correlation between folds.
}

\makeatletter%
\@ifclassloaded{imsart}{%
    \RequirePackage{amsthm,amsmath,amsfonts,amssymb}
    \RequirePackage[authoryear]{natbib}
    \RequirePackage[colorlinks,citecolor=blue,urlcolor=blue]{hyperref}
    \RequirePackage{graphicx}

    \startlocaldefs
    \theoremstyle{plain}

    \newtheorem{theorem}{Theorem}[section]
    \newtheorem{lemma}[theorem]{Lemma}
    \newtheorem{corollary}[theorem]{Corollary}

    \theoremstyle{definition}
    \newtheorem{definition}[theorem]{Definition}
    \newtheorem*{example}{Example}
    
    \newtheorem{assumption}{}
    \newtheorem{condition}{}
    \endlocaldefs

}{%

\@ifclassloaded{biom}{

  \makeatletter
  \def\proof{\@ifnextchar[{\@proof}{\@proof[Proof]}}
  \def\@proof[#1]{%
  \reset@font\rm \trivlist \item[\hskip \parindent%
  {\reset@font~~ \it #1.}]}

  \@namedef{proof*}{\@ifnextchar[{\@sproof}{\@sproof[Proof]}}
  \def\@sproof[#1]{%
  \reset@font\rm \trivlist \item[\hskip \labelsep%
  {\reset@font\it #1.}]}
  \@namedef{endproof*}{\endtrivlist}

  \makeatother

}{
  \usepackage{arxiv}
  \usepackage{standalone}
  \usepackage{multirow}
  \usepackage{setspace}
  \usepackage{amsthm}
  \usepackage[round]{natbib}
  \usepackage{svg}

  \newcommand{\authorlist}{
    Salvador V.~Balkus \\
    Department of Biostatistics,\\
    Harvard Chan School of Public Health,\\
    \texttt{sbalkus@g.harvard.edu}\\
    \And
    Hasan Laith \\
    Department of Statistics,\\
    Harvard College,\\
    \texttt{hasanhussein@college.harvard.edu}\\
    \And
    Nima S.~Hejazi \\
    Department of Biostatistics,\\
    Harvard Chan School of Public Health,\\
    \texttt{nhejazi@hsph.harvard.edu}\\
  }

  \newtheorem{theorem}{Theorem}
  \newtheorem{lemma}{Lemma}
  
  \newtheorem{definition}{Definition}
  \newtheorem{corollary}[theorem]{Corollary}
  {\theoremstyle{definition}\newtheorem{assumption}{}}
  {\theoremstyle{definition}}
  {\theoremstyle{definition}}
}

}
\makeatother

\newcommand{\E}{\mathbb{E}}

\renewcommand{\P}{\mathsf{P}}
\newcommand{\Q}{\mathsf{Q}}

\renewcommand{\Pr}{\mathbb{P}}
\newcommand{\Qr}{\mathbb{Q}}

\newcommand{\ind}{\mbox{$\perp\!\!\!\perp$}}

\newcommand{\hf}{\hat{f}}
\newcommand{\f}{f}
\newcommand{\hfe}{f_{\hat{\eta}}}

\newcommand{\St}{S_{\text{train}}}
\newcommand{\Se}{S_{\text{eval}}}
\newcommand{\Xt}{X_{\text{train}}}
\newcommand{\Xe}{X_{\text{eval}}}
\newcommand{\nt}{n_{\text{train}}}
\renewcommand{\ne}{n_{\text{eval}}}
\newcommand{\Pt}{\P_{\text{train}}}
\newcommand{\Pe}{\P_{\text{eval}}}
\newcommand{\Pet}{\P_{\text{eval}, \text{train}}}
\newcommand{\Pect}{\P_{\text{eval} \mid \text{train}}}

\newcommand{\Var}{\text{Var}}

\makeatletter
\@ifclassloaded{biom}{%

\title[\titlepaper]{\titlepaper}

\author{Salvador V. Balkus$^{1,*}$\email{sbalkus@g.harvard.edu}, 
Hasan Laith$^{2,**}$\email{hasanhussein@college.harvard.edu}, and 
Nima S. Hejazi$^{1,***}$\email{nhejazi@hsph.harvard.edu} \\
$^{1}$Department of Biostatistics, Harvard T.H. Chan School of Public Health, Boston, MA 02115, USA \\
$^{2}$Department of Statistics, Harvard College, Cambridge, MA 02138, USA\\}

}{%
  \author{\authorlist}
  \date{\today}
}

\begin{document}

\@ifclassloaded{imsart}{
    \begin{frontmatter}
    \title{\titlepaper}
    \runtitle{On the use of cross-fitting with correlated units}

    \begin{aug}
    \author[A]{\fnms{Salvador}~\snm{Balkus}\ead[label=e1]{sbalkus@g.harvard.edu}\orcid{0000-0003-4695-833X}},
    \author[B]{\fnms{Hasan}~\snm{Laith}\ead[label=e2]{hasanhussein@college.harvard.edu}\orcid{0009-0007-9082-1564}}
    \and
    \author[C]{\fnms{Nima}~\snm{Hejazi}\ead[label=e3]{nhejazi@hsph.harvard.edu}\orcid{0000-0002-7127-2789}}

    \address[A]{Salvador Balkus is PhD Candidate, Department of Biostatistics,
    Harvard Chan School of Public Health, Boston, MA,
    USA\printead[presep={\ }]{e1}.}

    \address[B]{Hasan Laith is Undergraduate, Department of Statistics, Harvard
    College, Cambridge, MA, USA\printead[presep={\ }]{e2}.}

    \address[C]{Nima Hejazi is Assistant Professor, Department of
    Biostatistics, Harvard Chan School of Public Health, Boston, MA,
    USA\printead[presep={\ }]{e3}.}

    \end{aug}

    \begin{abstract}
    \abstracttext
    \end{abstract}

    \end{frontmatter}
}{

\@ifclassloaded{biom}{

\date{{\it Received May} 2026. {\it Revised May} 2026.  {\it
Accepted May} 2026.}



\pagerange{\pageref{firstpage}--\pageref{lastpage}} 
\volume{64}
\pubyear{2026}
\artmonth{May}


\doi{10.1111/j.1541-0420.2005.00454.x}


\label{firstpage}


\begin{abstract}
\abstracttext
\end{abstract}

%



\maketitle

}{
  \title{\titlepaper}
  \maketitle
  \makeatother
  \begin{abstract}
    \abstracttext
  \end{abstract}

}
}

\section{Introduction}

\begin{figure}[t]
\centering
\includegraphics[width=0.50\linewidth]{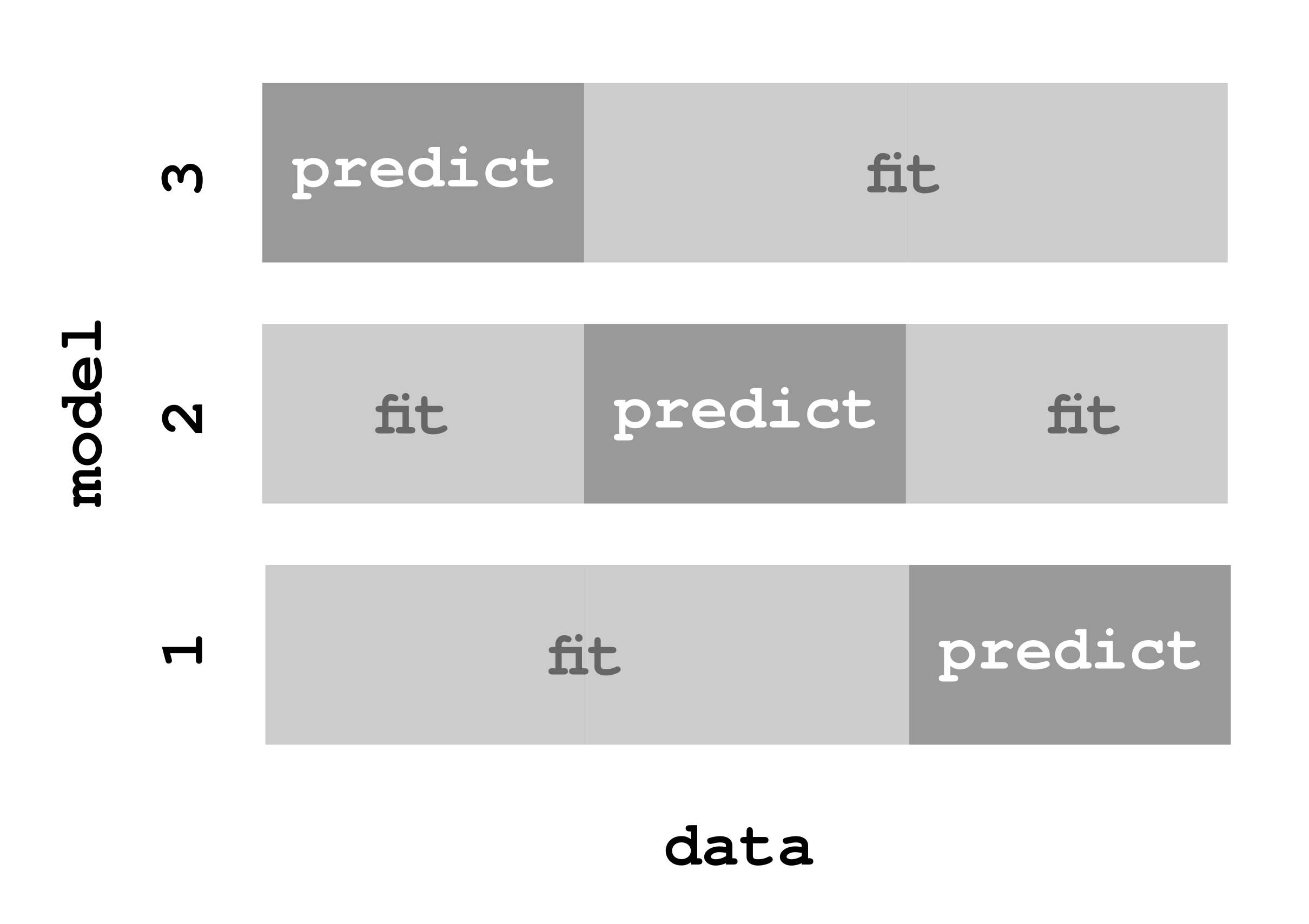}
\caption{As-independent cross-fitting of three models across three folds.}\label{fig:as-independent}
\end{figure}

Sample-splitting is a popular technique in statistical and machine learning, for
purposes ranging from optimal hyperparameter tuning and model selection to formal asymptotic analysis of statistical estimators. In any of its forms, sample-splitting proceeds by partitioning the available data into two distinct parts: one on which
an algorithm is fitted, termed the training or learning
fold, and another in which the trained algorithm is used to obtain predictions, termed the holdout or validation fold. \textit{Cross-fitting} applies sample-splitting repeatedly to ensure efficient use of every observation: data are split into several
folds, with prediction algorithms trained on all but one of the folds and used
to obtain predictions on the holdout fold (see Figure
\ref{fig:as-independent}).

Sample-splitting and cross-fitting have a long history in statistics, being discussed as
early as, for instance, \cite{mosteller1968data}, \cite{Stone1974}, and
\cite{Stone1978}, in the context of bias reduction and hyperparameter selection. Foundational work on semi-parametric methods relied on cross-fitting for the
elimination of key asymptotic error terms~\citep{Bickel1982, Bickel1988}.
\textit{Cross-validation}---the use of cross-fitting to
assess out-of-sample estimation error---grew immensely popular for the
evaluation of prediction algorithms in machine learning~\citep{picard1990data,
shao1993linear, Zhang1993}, especially with the introduction of formal
theory on the optimality of cross-validation-based selection~\citep{Dudoit2005,
van2006estimating}.

\begin{figure}[b]
\centering
\includegraphics[width=0.45\linewidth]{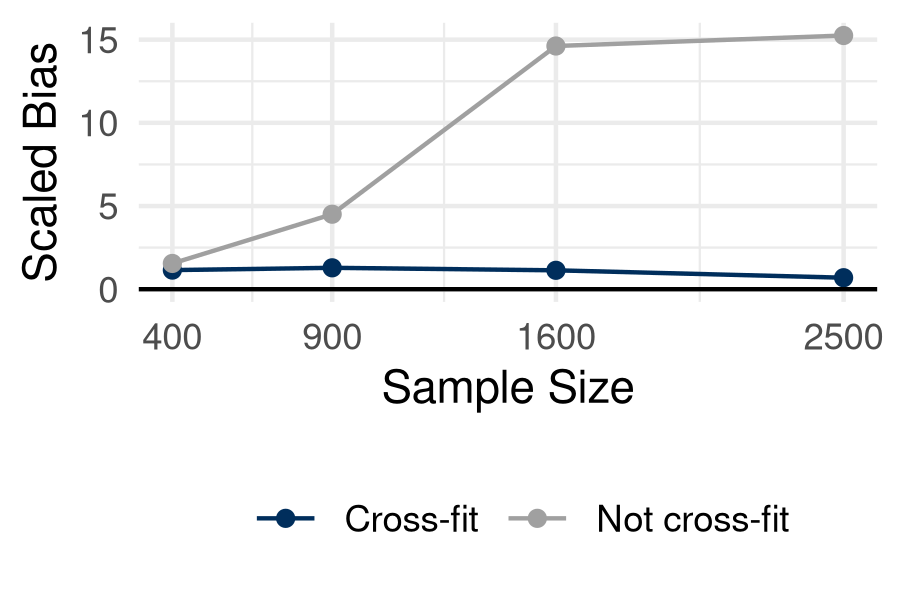}
\caption{Simulation of one-step ATE estimation using neural networks to learn the propensity score and outcome regression. Without cross-fitting, the bias fails to shrink fast enough for valid inference. 
}\label{fig:bias-elimination}
\end{figure}

Today, cross-fitting has grown especially popular in the field of \textit{causal machine learning}---the use of machine learning to perform statistical estimation and inference on causal effects. In this field, which encompasses targeted~\citep{van2011targeted, van2018targeted} and debiased \citep{Chernozhukov2018,
newey2018cross} machine learning, practitioners use cross-fitting to eliminate a critical
\textit{empirical process term}, a source of bias that arises in the asymptotic
analysis of semi-parametric estimators that rely on machine learning to estimate nuisance parameters. The use of cross-fitting avoids the need to otherwise impose specific structural assumptions on the function classes to which nuisance functions belong (e.g., Donsker conditions). Figure \ref{fig:bias-elimination} demonstrates an empirical example of this bias arising in a causal machine learning simulation that uses neural networks to estimate nuisance functions. One may note that this goal differs substantially from that of cross-validation, which concerns itself with evaluating how well a model performs, rather than eliminating key bias terms. The ubiquity of cross-validation in the arena of model evaluation, however, has led many to conflate the two procedures, especially in the context of correlated data. 

Foundational works in the causal machine learning literature often assume that study
units in a given dataset arise as independent observations. As the popularity of such methods has grown, a recently
growing strand of the literature has begun to extend such techniques to
settings where units may be correlated, such as clustered
data~\citep{Chiang2021, Fuhr2024}, spatial or network data~\citep{ogburn2022, Balkus2024,
Emmenegger2025}, and
time series~\citep{Semenova2023, vanderlaan2018}. In this context, many authors
have imported intuition---based on formal results---from the setting of
cross-validation to that of cross-fitting in order to extend the technique to the correlated data setting. We examine whether such extensions are necessary and justified.

Arguments in this vein often begin with the assumption that the purpose of
cross-fitting is to ensure that the study units used to train a prediction
algorithm are independent of those used to generate predictions. Consequently,
such works propose special cross-fitting schemes designed to keep correlated
units within the same training fold. Such arguments are justified when
estimating out-of-sample error; for instance, \cite{Rabinowicz2020} show that
when data units are not independent, a cross-validated estimate of mean squared
error (MSE) will be biased, and therefore requires correction. However, as we
demonstrate, such arguments do not extend to the standard setting of causal
machine learning, where cross-fitting is primarily used to eliminate empirical
process bias terms, not estimate MSE.

Why does this discrepancy exist? Unlike using cross-validation to estimate an out-of-sample MSE, which can be decomposed into both bias \textit{and} variance terms \citep{hastie2009elements}, cross-fitting involves the consistent estimation of \textit{only} a bias: the empirical process bias. Even for correlated units, the sample mean is an unbiased estimator; so, as long as the amount of dependence between units is weak enough for the estimator's variance to shrink to zero asymptotically, the empirical process bias will still be asymptotically negligible. 

So how should cross-fitting for causal machine learning be performed when study
units may be correlated? We argue that, in many correlated data settings, one
can use the same cross-fitting techniques as would be appropriate as if units
were independent. That is, \textit{ignoring correlation and simply splitting
units independently} is sufficient to eliminate empirical process bias. 

\textbf{Contribution}. In this manuscript, we establish a unifying theorem to justify the use of ``as-independent'' cross-fitting even in the correlated setting. First, Section \ref{sec:methods} formalizes our argument mathematically, demonstrating the precise conditions under which we can expect that the empirical process bias term will be asymptotically negligible, even when the correlation structure is ignored in the splitting process. Next, we use simulation experiments to demonstrate in Section \ref{sec:applications} that in many correlated data settings, even in finite samples, the performance of this na\"{i}ve technique
match---and, occasionally,
outperform---proposed techniques that eliminate correlation between folds. Section \ref{sec:conclusion} synthesizes our findings with a few concluding remarks. 

\section{Cross-fitting theory}\label{sec:methods}

In this section, we formally argue that even when correlation exists between
units in the observed dataset, ``as-independent'' cross-fitting will still
eliminate bias from the empirical process term of an estimating function-based
estimator, so long as the number of correlated units is not too large.

\subsection{Setup}

Throughout the manuscript, we will consider a vector of random variables $X = [X_1, \ldots, X_n]$ taking values on a common measurable space with support $\mathcal{X}$. The goal will be to estimate the expected value of some function $f$ applied to the data; if $X_i$ are stationary, meaning the marginal distributions $X_i \sim \P$ are identical, this is $\E[f(X_i)]$; otherwise, it is $n^{-1}\sum_{i=1}^n \E[f(X_i)]$. Typically, neither the true data distribution nor $f$ are known. Therefore, we want to perform estimation and inference by first estimating the function $\hat{f}$ and then taking its empirical mean $n^{-1}\sum_{i=1}^n \hat{f}(X_i)$.

In existing works, where $X_i \overset{\text{iid}}{\sim}\P$ , authors often adopt ``empirical process notation'', defining the population measure $\P f = \int f(x)d\P$, the expected value over the marginal distribution of $X_i$, and the empirical measure $\P_nf = n^{-1}\sum_{i=1}^n f(X_i)$.
To understand how the estimator $\P_n \hat{f}$ behaves after plugging in some arbitrary $\hat{f}$, they decompose its error into three terms:
\begin{align}\label{eq:decomposition}
    \P_n \hat{f} - \P f =&  \underbrace{(\P_n - \P) f}_{\text{CLT term}} +
    \underbrace{\P (\hat{f} - f)}_{\text{nuisance bias}}
    + \underbrace{(\P_n - \P)(\hat{f} - f)}_{\text{empirical
    process bias}} \ .
\end{align}
\noindent Each of the three terms above in Equation~\ref{eq:decomposition} plays a specific role in defining the
properties of the estimator $\P_n \hat{f}$:

\begin{enumerate}
    \item The \textbf{CLT} term, when appropriately scaled, will typically be asymptotically normal by some form of
      the Central Limit Theorem. When data units are independent, the classical CLT tells us that such convergence in distribution holds with scaling factor $\sqrt{n}$, permitting statistical inference based on the normal distribution.
      When $X_1, \ldots, X_n$ are correlated, one typically will rely on a
      \textit{dependent-data} CLT imposing assumptions about the nature of the
      dependence, one in which asymptotic normality may be achieved after scaling by a possibly slower rate $r_n$. 
    \item The \textbf{nuisance bias} term, also known as the second-order remainder, is controlled by estimation error of
      $\hat{f}$, which usually depends on the form of $f$. Commonly, $f$ is indexed by a set of nuisance parameters $\eta$, each of which may be estimated such that $\hat{f} = f_{\hat{\eta}}$. If we choose a set
      of algorithms to fit the combination of nuisance models such that each
      function in $\hat{\eta}$ converges to its corresponding true function
      $\eta$ at a fast enough asymptotic rate, then this term vanishes. For
      instance, for parametric $\hat{\eta}$, this rate will typically be
      attained at $o_{\P}(n^{-1/2})$; for $f$ chosen to be an
      influence function
      with doubly-robust second-order remainder 
      (see Equation~\ref{eq:causal-eif} later on), then machine learning
      algorithms whose rate \textit{product} matches $o_{\P}(n^{-1/2})$ may be
      employed. See \cite{van2011targeted} or \cite{Chernozhukov2018} for more
      details.
    \item The \textbf{empirical process} term can be thought of as the error
      from evaluating $f_{\hat{\eta}}$ on just a sample of data, rather than
      the whole population. Bounding this term can be thought of intuitively as
      answering the question ``\textit{if one had the true $f_{\eta_0}$, could
      they consistently estimate the bias $\P(f_{\hat{\eta}} - f_{\eta_0})$ using
      only the sample observed?}''
\end{enumerate}

In causal machine learning (e.g., targeted minimum loss estimation,
double machine learning), cross-fitting ensures the empirical
process term $(\P_n - \P)(f_{\eta} - f_{\hat{\eta}})$ converges to zero
(i.e., ``vanishes'') at a fast enough rate so that the CLT may be used to conduct statistical inference. An alternative approach is to assume that $f_{\hat{\eta}}$ lies in a
\textit{Donsker class}, a particular function class that restricts $f$ (or,
equivalently $\eta$) such that the empirical process term must achieve the
desired rate. \cite{bickel1993efficient}, \cite{van2000asymptotic}, and
\cite{kosorok2008introduction} discuss these classes in depth; for example, see Lemma 19.24 in \cite{van2000asymptotic}. For example,
most parametric models fall into a Donsker class. 

However, it has been known that in the setting of independent units, the use of
cross-fitting ``buys out'' the need for such assumptions by eliminating this
term for \textit{any} nuisance estimator $\hat{\eta}$ \citep{Schick1986, klaassen1987consistent, newey2018cross, Chernozhukov2018}. As specifically noted by \cite{newey2018cross}, when $f_{\hat{\eta}}$ is evaluated on the same data upon which it is trained, its residuals may suffer an optimistic ``same-observation bias''. Splitting the data ensures that the empirical process term is an unbiased estimator of zero. 

As an example, consider data $X = (L, A, Y)$, where $Y$ represents an outcome of interest, $A$ some treatment or exposure of interest, and $L$ a set of confounders. In causal inference one often estimates the counterfactual mean $\E(\mu(a, L))$, where $\mu$ denotes a conditional mean functional $\mu(A, L) = E(Y \mid A, L)$. In this setting, a common estimator is the doubly-robust one-step estimator
\begin{equation*}\label{eq:causal-eif}
    \P_n f_{\hat{\eta}} = \frac{1}{n}\sum_{i=1}^n \Big(\hat{\mu}(a, L_i) + \frac{I(A_i = a)}{\hat{g}(a \mid L_i)}(Y_i - \hat{\mu}(A_i, L_i))\Big) \ .
\end{equation*}
Here, the nuisance parameters are $\eta = (\mu, g)$, where $g$ represents the density of the treatment. Often, one desires to learn these using some machine learning algorithm. Since such algorithms may predict much more effectively on data which they have ``seen'' compared to data which they have not, cross-fitting becomes a useful way to eliminate the empirical process bias arising from such an issue. 

When units are not independent and identically distributed, however, this notation and decomposition can hide crucial details, especially when sample splitting is involved. Therefore, we will require more specific notation. Let $\Pt$ denote the marginal distribution of the training data upon which the nuisances within $\hat{f}$ are trained, and $\Pe$ the marginal distribution of the evaluation data upon which predictions $\hf(X_i)$ and $\f(X_i)$ are produced. Furthermore, let $\P = \Pet$ denote their joint distribution, and $\Pect$ the distribution of the evaluation data conditional on the training data. Then, an equivalent decomposition can be stated by instead adding and subtract integrals over $\Pect$, like so:

\begin{align}\label{eq:decomposition2}
    \P_n \hat{f} - \P f =&  \underbrace{(\P_n - \P) f}_{\text{CLT term}} +
    \underbrace{\Pect (\hat{f} - f)}_{\text{nuisance bias}}
    + \underbrace{(\P_n - \Pect)(\hat{f} - f)}_{\text{empirical
    process bias}} \ .
\end{align}

In the classical decomposition with IID data, $\P = \Pe = \Pect$, so authors often interchange them, or exploit the law of total expectation to reason about $\P \hat{f}$ when the training data is fixed---see, for example, Lemma 1 of \cite{kennedy2024semiparametric} or Chapter 4.1 of \cite{schuler2024introduction}. But when units may be correlated, $\P \neq \Pe \neq \Pect$. Centering the empirical process term with $\Pect$, the marginal distribution, makes the empirical process term mean-zero under the true measure of the observed data. 

Many existing works seek to, in the actual data analysis, construct a sample-split such that $\Pet \approx \Pe \times \Pt$,  so that arguments proceed identically to the classical setting, but doing so can be challenging, computationally intensive, or involve discarding a significant number of units. With this slight modification to the decomposition in hand, we will proceed to an asymptotic analysis that shows such procedures are typically unnecessary. 

\subsection{Main Theorem}

With an appropriate empirical process term for dependent units defined, let us formally define the ``as-IID
sample-splitting'' procedure:

\begin{definition}[As-independent sample splitting]\label{def:as-iid}

Choose an integer $k$. Let $\St$ be a random sample of size $\lfloor
n(1 - 1/k)\rceil$ sampled without replacement from the set of unit
indices $I_n$. Let $\Se = I_n - \St$ containing $N \approx n/k$ units. Finally, let $\Xt = \{X_i : i \in \St\}$ and $\Xe = \{X_i : i \in \Se\}$. This
produces a ``split sample'' $\Xt$ and $\Xe$ that partitions the set of unit
indices $I_n$. We term this procedure ``as-independent'' sample-splitting.

\end{definition}

Next, let $\hf = \hfe$ denote an estimating function indexed by nuisance parameters $\hat{\eta}$ trained on
$\Xt$, and for shorthand, denote the empirical process term by 
\begin{equation*}
    Z_n = (\P_{\ne} - \Pect)(\hf - \f) = \frac{1}{\ne}\sum_{i \in \Se}\Big(\hf(X_i) - \f(X_i) - \E \Big[\hf(X_i) - \f(X_i) \mid \Xt\Big]\Big)
\end{equation*}

Before presenting the main result, we define two important assumptions needed for it to hold. 

\begin{assumption}[Central limit theorem]\label{assume:clt}
    For some $r_n \in [1, \sqrt{n}]$, a CLT exists over the dependent data. Specifically, for a fixed function $g$, denote
    \begin{equation*}
        \sigma_n^2(g) = \frac{r_n^2}{n^2}\sum_{i,j}\text{Cov}[g(X_i), g(X_j)]
    \end{equation*}
    which from the covariance structure of the data we will assume satisfies $\sigma_n^2(g) = O(r_n\Pe g^2/\sqrt{n})$. Then,
    \begin{equation*}
        r_n(\P_n - \Pe)g/\sigma_n(g) \overset{\Pe}{\rightsquigarrow} \mathcal{N}(0, 1)
    \end{equation*}
    
\end{assumption}

Assumption \ref{assume:clt} implies a CLT holds under the correlation structure of the units in the evaluation set, an assumption usually required to perform statistical inference based on an asymptotic normal approximation. In the IID setting, or settings with a constant number of correlated units, $r_n = \sqrt{n}$, but this rate may be slower when the amount of correlation grows with $n$. 

\begin{assumption}[Mean squared error convergence]\label{assume:mse}
     $\Pet(\hf - \f)^2 = o(1/r_n)$; that is, the mean-squared error of $\hf$ as an estimator of $\f$ converges to zero faster than $1/r_n$. 
\end{assumption}

Assumption \ref{assume:mse} is satisfied, for example, if $\Pet(\hf - \f) = O(1/r_n^2)$, which for the common rate $r_n = \sqrt{n}$ is often attainable in parametric models. If $\hat{f}$ is a doubly-robust functional, as mentioned above, only the products of the nuisance norm must attain this rate; $O(1/\sqrt{n})$ is attainable by many common statistical algorithms \citep{gyorfi2002distribution} in IID data. Furthermore, under our decomposition, controlling the nuisance bias requires  $[\Pect (\hf - \f)^2]^{1/2} = o_{\Pt}(1/r_n)$ anyways; if this holds, then Assumption \ref{assume:mse} will also hold as long as $r_n[\Pect (\hf - \f)^2]^{1/2}$ is uniformly integrable \cite[pg. 361]{billingsley2013convergence}.

\begin{assumption}[Mutual absolute continuity]\label{assume:absolute-continuity}
    $\Pe \times \Pt$ and $\Pet$ are mutually absolutely continous; that is, for some event $A$, $\Pe(A)\cdot \Pt(A) = 0 \iff \Pet(A) = 0$.
\end{assumption}

Assumption \ref{assume:absolute-continuity} is a technical condition required to ensure the Radon-Nikodym derivative $d\Pet / d(\Pe \times \Pt)$ exists in subsequent proofs; it holds if, for example, the densities of $\Xt$ and $\Xe$ exist with respect to some dominating measure, such as the Lebesgue measure \cite[pg. 449]{billingsley2013convergence}. It fails if, for example, the dataset contains units $X_i, X_j$ that may be perfectly correlated. 

Before presenting the main theorem, we will prove two required intermediary lemmas.

\begin{lemma}[Joint convergence implies marginal convergence]\label{lemma:marginal-convergence}
   Consider two, possibly correlated random vectors $V$ and $W$ with identical supports whose product of marginals $\P_V\times\P_W$ is absolutely continuous with respect to the joint probability $P_{V, W}$. If, for a nonnegative function $g_n$,  $\int g_n(v, w) d\P_{V, W} = o(1)$, then 
    \begin{equation*}
            \int g_n(v, w) d\P_{V} = o_{\P_{W}}(1)\ .
    \end{equation*}
\end{lemma}

\begin{proof}
    We will first prove that $\int g_n(v, w) d\P_{V, W} = o(1) \implies \int g_n(x, y) d(\P_{V} \times\P_{W}) = o(1)$, and then apply Markov's inequality to show convergence in probability. 
    
    Let $\rho = d(\P_V \times \P_W) / d\P_{V, W}$. By absolute continuity, we can split the integral over the  over the joint density into two parts by multiplying and dividing by the joint density:
    \begin{align*}
        \int g_n(v, w) d(\P_{V} \times \P_{W}) = \int g_n(v, w)\rho d\P_{V, W} &= \int_{\rho \leq M} g_n(v, w)\rho d\P_{V, W} + \int_{\rho > M} g_n(v, w)\rho d\P_{V, W}\\
        &\leq M\int_{\rho \leq M} g_n(v, w) d\P_{V, W} + \int_{\rho > M} g_n(v, w)\rho d\P_{V, W}
    \end{align*}
    We need that, for some $n$ large enough, both terms are bounded by $\varepsilon/2$. Choose $M$ such that $\int_{\rho > M} g_n(v, w)\rho d\P_{V, W} < \varepsilon/2$. Since $\int g_n(v, w) d\P_{V, W} = o(1)$, this means we can choose $n \geq N$ such that $\sup_{n \geq N}\int g_n(v, w) d\P_{V, W} \leq \delta$ for any positive $\delta$. Therefore, choose $\delta = \varepsilon/2M$. In this case, we have
    \begin{equation*}
        M\int_{\rho \leq M} g_n(v, w) d\P_{V, W} + \int_{\rho > M} g_n(v, w)\rho d\P_{V, W} \leq \frac{\varepsilon}{2} + \frac{\varepsilon}{2} = \varepsilon
    \end{equation*}
    Since it is possible to choose an $n$ large enough so that $\int g_n(v, w) d(\P_{V}\times \P_{W}) \leq \varepsilon$ for any $\varepsilon > 0$, it follows that $\int g_n(v, w) d\P_{V, W} = o(1) \implies \int g_n(v, w) d(\P_{V}\times \P_{W}) = o(1)$.

    Finally, combining this with Markov's inequality,
    \begin{equation*}
        \P_{W}\Big(\int g_n(v, w) d\P_V > t\Big) \leq \frac{\int g_n(v, w)d(\P_V \times \P_W)}{t} = o(1)\ ,
    \end{equation*}
    meaning that $\int g_n(x, y) d\P_{X} = o_{\P_{Y}}(1)$ as desired.
\end{proof}

\begin{lemma}[Information-theoretic bound between probability measures.]\label{lemma:kl-divergence}

Let $\P$ and $\Q$ denote arbitrary probability distributions, with $\P$ absolutely continuous with respect to $\Q$. Suppose that $g(W)$ for $W \sim \Q$ is a sub-exponential random variable, meaning that $\Q e^{\lambda(g - \Q g)} \leq e^{\nu^2 \lambda^2}$ for $-1/\nu \leq \lambda \leq 1/\nu$. Then,
\begin{equation*}
    |(\P - \Q) g |\leq 2\nu \sqrt{D_{\text{KL}}(\P \parallel \Q)}
\end{equation*}
\end{lemma}

This lemma tells us that the difference between an existing probability measure $\P$ and an ``ideal'' measure $\Q$ is bounded by the product of the variance parameter of the random variable under the ideal $\Q$ and the KL-divergence between $\P$ and $\Q$. It originates from \cite{xu2017information}, who proved it for sub-Gaussian random variables. We verify it also holds for sub-exponential variables using the same steps, reproduced here for convenience. 

\begin{proof}
The Donsker-Varadhan variational representation states that, for functions $g$ satisfying $\Q(e^g) \leq \infty$,

\begin{equation*}
    D_{\text{KL}}(\P \parallel \Q) = \sup_{g} \Big(\int g d\P - \log \int e^g dQ\Big)
\end{equation*}

Consequently, applying this to $W \sim \P$ and $W' \sim \Q$, scaled by $\lambda \in [-1/\nu, 1/\nu]$,
\begin{align*}
    D_{\text{KL}}(\P \parallel \Q) &\geq \P(\lambda g) - \log \Q(e^{\lambda g})\\
    &\geq \P(\lambda g) - \log \Q(e^{\lambda \Q g}) - \log \Q(e^{\lambda (g - \Q g)}) \tag{Add and subtract $\lambda\Q g$}\\
    &\geq \lambda (\P - \Q)g - \log \Q(e^{\lambda (g - \Q g)}) \tag{Jensen's inequality}\\
    &\geq \lambda (\P - \Q)g - \nu^2 \lambda^2 \tag{$W'$ is sub-exponential}
\end{align*}
Now, let $a = \nu^2$, $b = \lambda (\P - \Q)g$, and $c = D_{\text{KL}}(\P \parallel \Q)$. Rerranging algebraically, we have $a\lambda^2 - b\lambda + c \geq 0$. Since this is a positive quadratic function in $\lambda$, its discriminant $\sqrt{b^2 - 4ac}$ must be positive, meaning that $|b| \leq 2ac$; that is,
\begin{equation*}
    |(\P - \Q)g| \leq 2\nu \sqrt{D_{\text{KL}}(\P \parallel \Q)}
\end{equation*}
\end{proof}
Next, we present the main theorem. 

\begin{theorem}[As-independent sample-splitting in correlated data.] \label{thm:as-ind-ss}

Under Assumptions \ref{assume:clt} and \ref{assume:mse}, and the as-IID sample split in Definition \ref{def:as-iid}, the
empirical process term satisfies
\begin{equation*}
  Z_n = o_{\P}(1/r_n)
\end{equation*}
even under possibly correlated data units, which is asymptotically negligible. 
\end{theorem}

In classical CLT arguments, $r_n = \sqrt{n}$, but this rate may be slower in settings with correlated data. Let us proceed with the proof. 

\begin{proof}
The strategy of the proof will be to decompose the behavior of the empirical process term into an ``ideal'' term, governed by independent training and evaluation datasets, and a difference that can be bounded by the mutual information between the datasets. 

Recall that we can express the distribution of our data as $\P = \Pet$, the joint distribution of the training and evaluation sets. Let $\Q = \Pt \cdot \Pe$, the product of the marginal distributions of the training and evaluation sets. $\Q$ represents our ``ideal'' measure where the training and evaluation datasets used in the sample splitting are independent. Also, let $\Pr(A)$ and $\Qr(A)$ denote the probabilities of an event $A$ according to the measure $\P$ and $\Q$, respectively. Finally, let $Z_n$ denote the empirical process term. Our goal will be to prove that 
\begin{equation*}
\Pr(|r_n Z_n| > \varepsilon) = o(1)
\end{equation*}

That is, as we increase the size of the training dataset and the size of the evaluation dataset, our sample-split empirical process term will converge to zero. 

We will begin by adding and subtracting our ideal measure; by the triangle inequality,
\begin{equation*}
    \Pr(|r_n Z_n| > \varepsilon) \leq |\Qr(|r_n Z_n| > \varepsilon)| + |\Pr(|r_n Z_n| > \varepsilon) - \Qr(|r_n Z_n| > \varepsilon)|
\end{equation*}

First, we will analyze $\Qr(|r_n Z_n| > \varepsilon)$. Under independence, conditioning on $\Xt$ does not change the distribution of $\Xe$, so we will exploit the law of total probability representation $\Qr(|r_n Z_n| > \varepsilon) = \E_{\Xt}[\Qr(|r_n Z_n| > \varepsilon \mid \Xt)]$. Then, using Assumption \ref{assume:clt} and Chebyshev's inequality, since $\Xt \ind \Xe$,
\begin{align*}
   \limsup_{\nt\rightarrow\infty} \limsup_{\ne\rightarrow\infty} \E_{\Xt}[\Qr(|r_n Z_n| > \varepsilon \mid \Xt)] 
    &=  \limsup_{\nt\rightarrow\infty}\E_{\Xt}\Big[\Qr(Z > \varepsilon \mid \Xt)\Big] \\
    &\leq\E_{\Xt}\Big[\limsup_{\nt\rightarrow\infty}\Qr(Z > \varepsilon \mid \Xt)\Big]
\end{align*}

where $Z\mid \Xt \sim N((\Pe - \Pect)(\hf - \f), \sigma^2(\hat{f} - f))$; here $\hf$ is fixed through the conditioning on the training data. Note that moving the limit to the inside of the expected value is permissable by the reverse Fatou lemma, since distribution functions are bounded by 1. Finally, note that 
\begin{equation*}
   \Pt(\Pe - \Pect)(\hf - \f) = (\Pe\times\Pt - \Pet)(\hf - \f) \leq [(\Pe\times\Pt - \Pet)(\hf - \f)^2]^{1/2} 
\end{equation*}
by the law of total probability followed by Jensen's inequality. This means
\begin{align*}
    \E\Big[\limsup_{\nt\rightarrow\infty}\Qr(Z > \varepsilon \mid \Xt)\Big] &= O([(\Pe\times\Pt - \Pet)(\hf - \f)^2]^{1/2}) + O(\sigma^2(\hat{f} - f))\\
    &= O([o(1/r_n)]^{1/2}) + O(o(1/r_n)) = o(1)\ . 
\end{align*}

All that remains is to show $|\Pr(|r_n Z_n| > \varepsilon) - |\Qr(|r_n Z_n| > \varepsilon)| = o(1)$: that the difference in the probability that $r_nZ_n > \varepsilon$ between the ``ideal'' and ``actual'' probability measures converges to 0. By the same logic we argued previously,
\begin{align*}
     &\limsup_{\nt\rightarrow\infty}\limsup_{\ne\rightarrow\infty}|\Pr(|r_n Z_n| > \varepsilon) - \Pt[\Qr(|r_n Z_n| > \varepsilon \mid \Xt)]| \\&=  \limsup_{\nt\rightarrow\infty}\limsup_{\ne\rightarrow\infty}|\Pr(|r_n Z_n| > \varepsilon) - \Pt[\Qr(Z > \varepsilon \mid \Xt)]|\\
    &\leq \limsup_{\nt\rightarrow\infty}\limsup_{\ne\rightarrow\infty}|\Pr(|r_n Z_n| > \varepsilon) - \Pt[\Qr(\sigma_n(\hf - \f)\mathcal{N}(0, 1) > \varepsilon \mid \Xt)]| \\
    &+ o\Big([(\Pe\times\Pt - \Pet)(\hf - \f)^2]^{1/2}\Big)
\end{align*}
We already showed $o\Big([(\Pe\times\Pt - \Pet)(\hf - \f)^2]^{1/2}\Big) = o(1)$ previously. Applying Chebyshev's inequality to the two components within the first term separately, the above is bounded by 
\begin{equation*}
    |\Pr(|r_n Z_n| > \varepsilon) - \Pt[\Qr(\sigma_n(\hf - \f)\mathcal{N}(0, 1) > \varepsilon \mid \Xt)]| \leq |\Pet(r_n Z_n)^2 + (\Pt\times\Pe)[\sigma_n^2(\hf - \f)\cdot \chi(1)]|
\end{equation*}

Now, $\sigma_n^2(\hf - \f)\cdot \chi(1)$ is sub-exponential with parameter $\nu = \sigma_n^2(\hf - \f)$. Combining Assumption \ref{assume:mse} and Lemma \ref{lemma:marginal-convergence}, 
\begin{equation*}
    \Pt\sigma_n^2(\hf - \f) = O(r_n (\Pt\times \Pe) (\hf - \f)^2/\sqrt{n}) = o(1/\sqrt{n})
\end{equation*}

Therefore, leaning on Assumption \ref{assume:absolute-continuity}, Lemma \ref{lemma:kl-divergence} can be applied to see
\begin{align*}
    |\Pr(|r_n Z_n| > \varepsilon) - \Qr(|r_n Z_n| > \varepsilon)| &\leq \Pt\sigma_n^2(\hf - \f)\sqrt{D_{\text{KL}}(\Pet \parallel \Pe \times \Pt)}\\
    &= o(1/\sqrt{n})\sqrt{I(\Xt; \Xe)}
\end{align*}
where $I(\Xt; \Xe)$ denotes the mutual information between the training data and the evaluation data. 

Recall that the mutual information is bounded by joint entropy, which is bounded by the entropy of individual units. The entropy of a single random variable with variance $\sigma^2$ is bounded by the entropy of a $\mathcal{N}(0, \sigma^2)$; putting this together, we can write

\begin{equation*}
    I(\Xt; \Xe) \leq H(\Xt, \Xe) \leq \sum_{i=1}^n H(X_i) \leq n \max_i \log[2\pi e \Var(X_i)] = O(n)
\end{equation*}

The last equality follows from the assumption that $\max_i \Var(X_i)$ is bounded, which is required for the CLT Assumption \ref{assume:clt} to hold. Note that the above assumes $\Var(X_i) > 1/2\pi e$ to ensure the differential entropy is positive; one can always rescale the random variable if it not, in order to make this the case. Plugging this into our bound, we have that

\begin{equation*}
    |\Pr(|r_n Z_n| > \varepsilon) - \Qr(|r_n Z_n| > \varepsilon)| = o(1/\sqrt{n})\sqrt{O(n)} = o(1)\ ,
\end{equation*}
meaning that
\begin{equation*}
    \Pr(|r_n Z_n| > \varepsilon) \leq \Qr(|r_n Z_n| > \varepsilon) +  |\Pr(|r_n Z_n| > \varepsilon) - \Qr(|r_n Z_n| > \varepsilon)| = o(1) + o(1) = o(1)
\end{equation*}

which completes the proof of Theorem \ref{thm:as-ind-ss}. 

\end{proof}

As a final remark, we note that the ability of as-independent cross-fitting to
eliminate the empirical process term follows from the ability of
as-independent sample-splitting to do so. 

\section{Applications}\label{sec:applications}

Theorem~\ref{thm:as-ind-ss}
holds across many correlated-data settings. In this section, we
demonstrate how it
can be used to
show the negligibility of the empirical process term for causal machine
learning in three examples: clustered data, network data, and time series. Before we begin, we define a preliminary lemma to be employed repeatedly in the proofs of subsequent corollaries. 

\begin{lemma}[Bienaym\'{e} Bound]\label{lemma:bienayme}
    Suppose the random sample $X_1, \ldots, X_n$ contains fewer than $C_n$ pairs of correlated units. Then,
    \begin{equation}\label{eq:bienayme}
        \text{Var}\Big(r_n(\P_n - \P)(f_{\hat{\eta}_n} - f_{\eta_0})\Big) \leq \frac{r_n^2}{n^2}\cdot 2C_n \cdot  \P(f_{\hat{\eta}_n} - f_{\eta_0})^2
    \end{equation}
\end{lemma}

\begin{proof}
    Recall Bienaym\'{e}'s identity, which states that $\text{Var}\Big(\sum_{i=1}X_i\Big) = \sum_{i, j}\text{Cov}(X_i, X_j)$. Applying this to the Equation~\ref{eq:bienayme}, 
    \begin{align*}
        \text{Var}\Big(r_n(\P_n - \P)(f_{\hat{\eta}_n} - f_{\eta_0})\Big) &= \frac{r_n^2}{n^2} \cdot \sum_{i,j} \text{Cov}\Big((f_{\hat{\eta}_n} - f_{\eta_0})(X_i), (f_{\hat{\eta}_n} - f_{\eta_0})(X_j)\Big)\\
        &\leq \frac{r_n^2}{n^2} \cdot 2C_n \cdot \max_{i,j} \text{Cov}\Big((f_{\hat{\eta}_n} - f_{\eta_0})(X_i), (f_{\hat{\eta}_n} - f_{\eta_0})(X_j)\Big) \ .
    \end{align*}
    Note that $\text{Cov}(X_i, X_j) = \text{Corr}(X_i, X_j)\cdot \text{Var}({X_i}) \leq \text{Var}(X_i)$ using the assumption of stationary data and the fact that correlations are bounded in $[0, 1]$. Then, $\text{Var}\Big((f_{\hat{\eta}_n} - f_{\eta_0})(X_i)\Big) = \P(f_{\hat{\eta}_n} - f_{\eta_0})^2$; therefore,
    \begin{equation*}
        \frac{r_n^2}{n^2} \cdot 2C_n \cdot\max_{i,j} \text{Cov}\Big((f_{\hat{\eta}_n} - f_{\eta_0})(X_i), (f_{\hat{\eta}_n} - f_{\eta_0})(X_j)\Big) \leq \frac{r_n^2}{n^2}\cdot 2C_n \cdot \P(f_{\hat{\eta}_n} - f_{\eta_0})^2 \ ,
    \end{equation*}
which completes the proof. 
\end{proof}

\subsection{Clustered Data}

In clustered data (which also encompasses ``panel'', ``longitudinal'', and
``repeated measures'' data structures), each unit belongs to one or more
clusters. Units within the same cluster are considered to be correlated.
Causal machine learning procedures have been developed for this setting
by~\cite{Klosin2023}, \cite{Clarke2025}, and~\cite{Fuhr2024}, among others,
typically for a single cluster variable.

When using cross-fitting, authors often imply that special care must be taken to avoid correlation across folds. However, Theorem~\ref{thm:as-ind-ss}
contradicts this idea, suggesting little practical difference
between splitting independently (ignoring any clustering) and keeping units
from the same cluster within the same fold, at least asymptotically. Indeed,
simulation results from \cite{Fuhr2024} support this argument, demonstrating
that as-independent cross-fitting generally tends to be the same as, or even
outperforms, other methods.

\begin{figure}[!hb]
\centering
\includegraphics[width=0.5\linewidth]{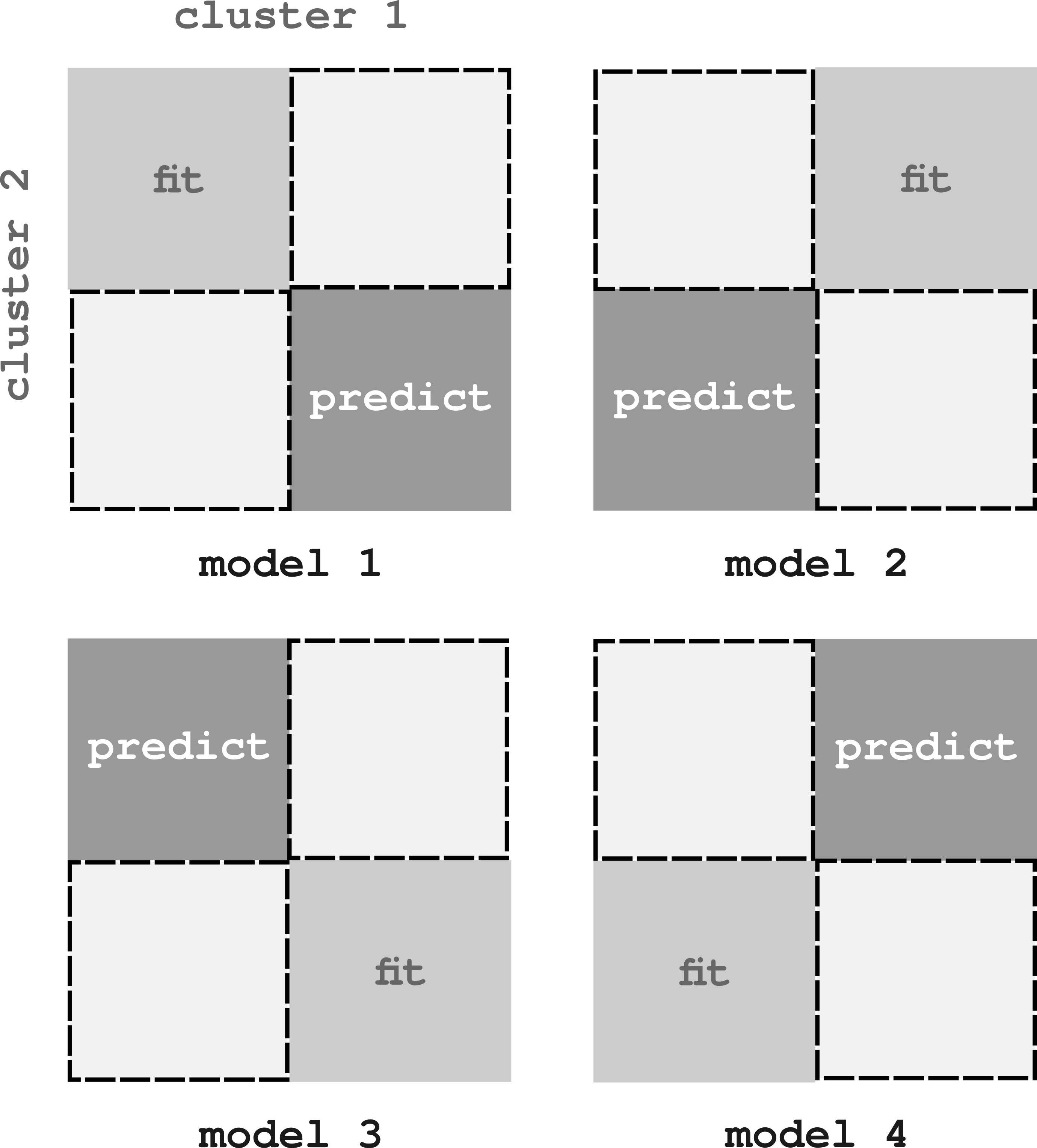}
\caption{Two-way cross-fitting introduced by  \protect\cite{Chiang2021}.}\label{fig:two-way}
\end{figure}

Building on the work of~\cite{Fuhr2024}, we investigate a slightly more
complicated data structure. Consider observations $\{X_{ij}\}_{i =1, \dots, N;
j = 1, \dots, M}$ that satisfy a two-way clustering:
$$
  X_{ij} \ind X_{i'j'} \quad \text{whenever } i \neq i' \text{ and }
  j \neq j' \ .
$$
\cite{Chiang2021} extends the double machine learning framework
of~\cite{Chernozhukov2018} to this multiple cluster setting. In doing so, they
introduce, for two-way clustering, a $K^2$-fold multi-way cross-fitting scheme
(or a $K^\ell$ scheme for higher dimensions) that holds out one block along
every clustering dimension when estimating nuisance functions
(see Figure~\ref{fig:two-way}).
This scheme uses the complementary block to estimate $\psi$, the parameter of
interest. However, as a corollary to our main theorem, we show that such a
scheme is unnecessary in clustered data.

\begin{corollary}[Cross-fitting in clustered data]\label{cor:as-ind-cf} Let $n \coloneqq NM$, with each cell $(i,j)$ a unit in the sample-splitting
procedure of Definition \ref{def:as-iid} (i.e., $I = \{(i,j): 1 \leq i \leq N, \, 1 \leq j
\leq M \}$). Suppose the data $X_1, \ldots, X_n$ are correlated according to a two-way clustering structure. Let Assumptions \ref{assume:clt} and \ref{assume:mse} hold under the following:
\begin{enumerate}
    \item Two-way CLT: $\sqrt{n}(\P_n - \P) f_{\eta_0} \overset{d}{\to}\mathcal N (0, \sigma^2)$; see Section 4.2
      of~\cite{Chiang2021} or Section 3 of~\cite{Davezies2018}.
    \item Bounded cluster sizes: there is a constant $C$ such that
      every row contains at most $C$ columns and every column contains at most
      $C$ rows.
\end{enumerate}
Then, a cross-fit causal machine learning estimator constructed using the as-IID sample-splitting of Definition \ref{def:as-iid} satisfies
\begin{equation*}
(\P_{\ne} - \Pe) (f_{\hat{\eta}_{S_1}} - f_{\eta_0}) = o_{\P}(1/\sqrt{n}) \ ,
\end{equation*}
meaning that the empirical process bias term is 
negligible under two-way clustering with
bounded clusters.
\end{corollary}

\begin{proof}

Each cell $(i,j)$ is correlated with at most $(C-1)+(C-1) \leq
2C$ other cells (those sharing its row and those sharing its column), so the
total number of correlated pairs is at most $2Cn = O(n),$ whereas the total
number of pairs is $n^2$. Apply Lemma \ref{lemma:bienayme} with $r_n = \sqrt{n}$ to see that 
\begin{align*}
    \text{Var} \Big(\sqrt{n_2}(\P_{\ne} - \Pe)(f_{\hat{\eta}_{S_1}} - f_{\eta_0})\Big)
    &\leq \frac{n_2}{n_2^2}\cdot O(n)\cdot \P(f_{\hat{\eta}_{S_1}} - f_{\eta_0})^2\\
      &\leq O\Big(\P(f_{\hat{\eta}_{S_1}} - f_{\eta_0})^2\Big) \ .
\end{align*}

This means the variance bound needed for the Theorem \ref{thm:as-ind-ss} CLT holds; therefore, the empirical process term must be $o_\P(1/\sqrt{n})$, which is
asymptotically negligible by the two-way CLT rate. 

\end{proof}

\begin{figure}[!t]
  \centering
  \includegraphics[width=0.5\linewidth]{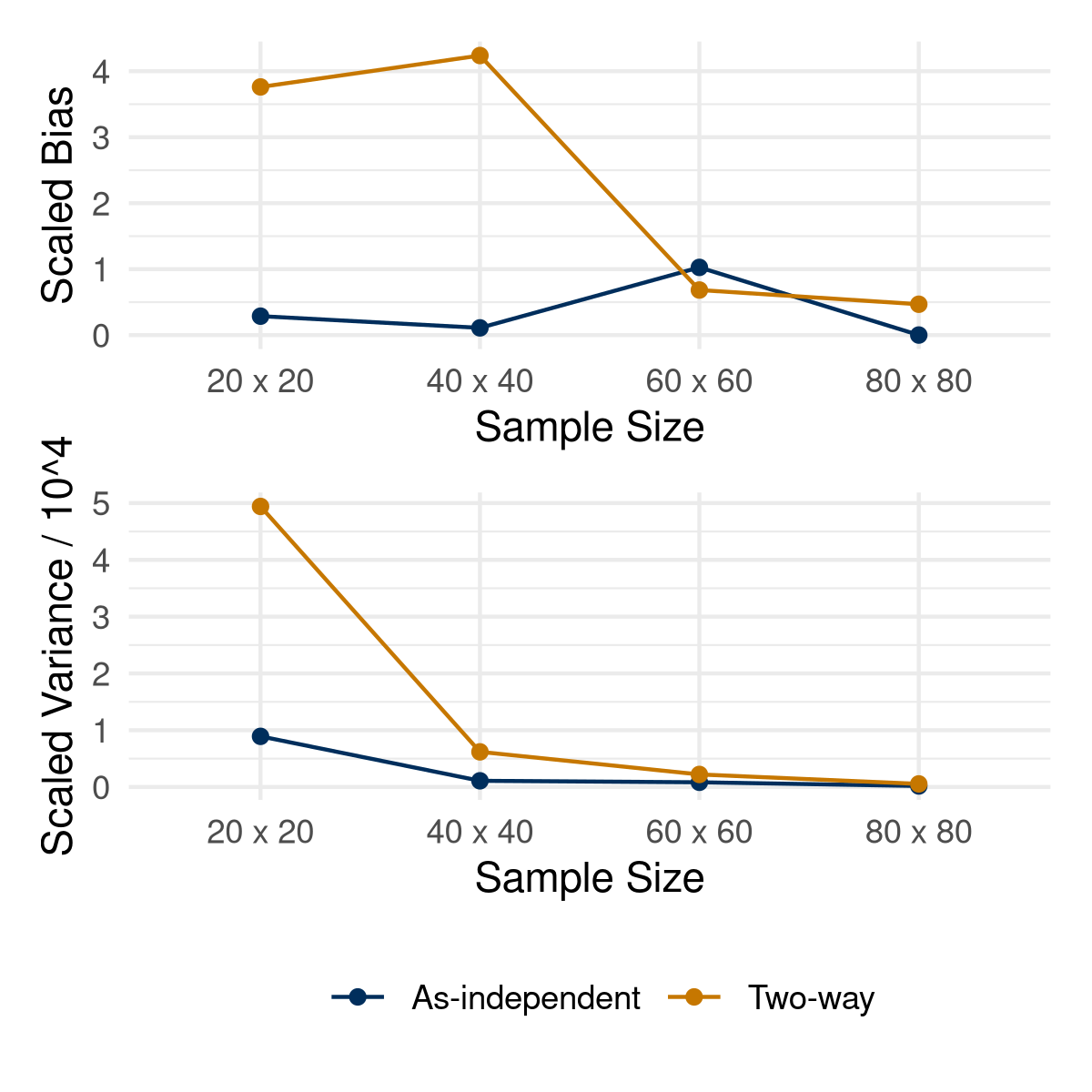}
  \caption{Operating characteristics of one-step ATE estimators in the cluster
  setting, comparing as-independent sample-splitting to the two-way method of \protect\cite{Chiang2021}.}\label{fig:cluster}
\end{figure}

\textbf{Simulation}. While our theoretical results indicate that the two-way
clustering scheme is unnecessary in this context, it may yet confer some
finite-sample benefits. To evaluate this, we compare the performance of as-IID
and two-way clustering~\citep{Chiang2021} cross-fitting on a two-way clustered
dataset with the following set-up:
\newline
\newline
\noindent \textit{Clustered errors}:
\begin{equation*}
    \gamma^{(L)}_i, \nu^{(L)}_j, \gamma^{(A)}_i, \nu^{(A)}_j, \gamma^{(Y)}_i,
    \nu^{(Y)}_j \sim \text{N}(0,1)
\end{equation*}
\textit{Covariates}:
\begin{align*}
    & L^{(1)}_{ij} \sim \text{N}(\gamma^{(L)}_i + \nu^{(L)}_j,1); L^{(2)}_{ij} \sim \text{N}(0,1)\\
    & L^{(3)}_{ij} \sim \text{N}(\sin(L^{(1)}_{ij}) + 0.3L^{(2)}_{ij}, 0.5)\\
    & L^{(4)}_{ij} \sim \text{N}(\mathbb{I}(L^{(1)}_{ij} > 0)L^{(2)}_{ij},0.5); L^{(5)}_{ij} \sim \text{N}(1,2)
\end{align*}
\textit{Treatment assignment mechanism}:
\begin{align*}
    A_{ij} \sim \text{Ber}\Big(\sigma(&-0.4 + 0.8L^{(1)}_{ij} -0.7(L^{(2)}_{ij})^2 +\\
    & 0.5\sin(L^{(3)}_{ij}) +
    0.4L_{ij}^{(1)}L_{ij}^{(2)} - \\ & 0.5L_{ij}^{(4)}L_{ij}^{(5)} +
    0.6\gamma^{(A)}_i - 0.6\nu^{(A)}_j)/5\Big)
\end{align*}
\textit{Outcome mechanism}:
\begin{align*}
    Y_{ij} \sim \text{N}\Big(&0.5L^{(1)}_{ij} - 0.4 L^{(2)}_{ij} + 0.3(L^{(3)}_{ij})^2 -
    0.5\sin(L^{(4)}_{ij})\\
    & + 0.4 L^{(1)}_{ij} L^{(2)}_{ij} +
    0.6\gamma^{(Y)}_i + 0.6\nu^{(Y)}_j + A_{ij}\\
    & + A_{ij}(0.5\sin(L^{(1)}_{ij}) - 0.5\sin(L^{(2)}_{ij})\\
    & + 0.3L^{(3)}_{ij}L^{(4)}_{ij} + 0.4\gamma_i^{\tau}\nu_j^{\tau})\Big)
\end{align*}
Across 500 replicates, we estimate the average treatment effect (ATE) using an
asymptotically efficient one-step estimator, with nuisance parameters estimated
using multivariate adaptive regression splines~\citep{Friedman1991} with the
\texttt{earth} package~\citep{earth} to learn the outcome and treatment
mechanisms. It is generally understood that data-adaptive machine learning algorithms like these do not fall into a Donsker class \citep{Smith2025}, necessitating cross-fitting. The results of our simulation experiment are shown in
Figure~\ref{fig:cluster}.

As Figure~\ref{fig:cluster}
shows, using two-way clustering decreases the estimator's empirical
performance fairly drastically at smaller sample sizes.
While as-independent cross-fitting keeps the estimator generally unbiased with
low variance, accounting for the correlation structure using two-way splitting
from \cite{Chiang2021} yields substantially increased bias and variance at
lower sample sizes. These differences disappear asymptotically.

\subsection{Network and Spatial Data}

Next, we consider data where observed units $X_i, X_j$ may be correlated if they share a neighbor in an
arbitrary undirected network with known adjacency matrix $G$. More formally,
\begin{equation*}
    X_i \ind X_j\quad \text{whenever } G_{ik} = 0 \text{ or } G_{jk} = 0, \forall k \in \{1, \ldots, n\}\ .
\end{equation*}

\begin{figure}[h]
\centering
\includegraphics[width=0.4\linewidth]{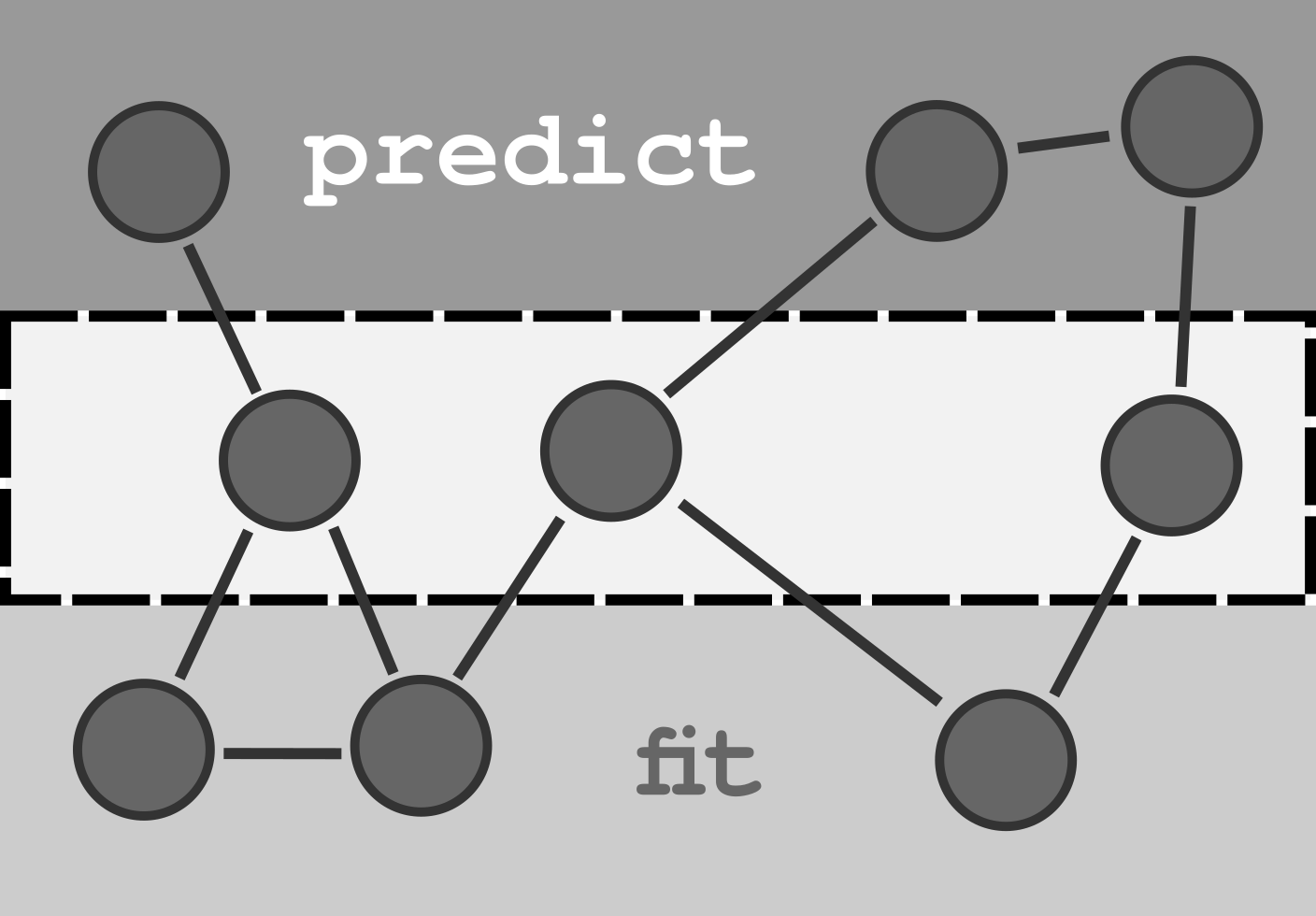}
\caption{Two-fold version of the cross-fitting method from \protect\cite{Emmenegger2025}.}\label{fig:network-folds}
\end{figure}

A few works consider causal machine learning in the network setting. 
\cite{Emmenegger2025} proposes a cross-fitting method that drops correlated
units from training folds (Figure~\ref{fig:network-folds}).
However, ~\cite{Balkus2024} have argued that as-independent cross-fitting
eliminates the empirical process term in this network scenario. We
prove Corollary~\ref{cor:as-ind-net}
by adapting a portion of the supplementary material of~\citet{Balkus2024}.

\begin{corollary}[as-IID cross-fitting in network data]\label{cor:as-ind-net}
Suppose the data $X_1, \ldots, X_n$ are correlated according to the aforementioned network structure $G$ with maximum node degree $K_n$. Let Assumptions \ref{assume:clt} and \ref{assume:mse} hold under the following:

\begin{enumerate}
    \item CLT for network-dependent data: $r_n(\P_n - \P)f_{\eta_0}
      \overset{d}{\rightarrow} \mathcal{N}(0, \sigma^2)$ for some $r_n \geq
      \sqrt{n}/K_n$; for example, see \cite{ogburn2022}.
    \item The maximum degree of $G$ satisfies the rate condition $K_n = o(\sqrt{n})$. 
\end{enumerate}

\noindent Then, a cross-fit causal machine learning estimator constructed using the as-IID sample-splitting of Definition \ref{def:as-iid} satisfies, for any given sample split,
\begin{equation*}
(\P_{\ne} - \Pe) (f_{\hat{\eta}_{S_1}} - f_{\eta_0}) = o_{\P}(1/r_n) \ ,
\end{equation*}
meaning that the empirical process bias term is 
negligible under network dependence with a ``slow-growing'' maximum degree.

\end{corollary}

\begin{proof}
    The number of correlated units is equivalent to the number of pairs $i, j$ satisfying, for some $k$, $G_{ik} \neq 0$ and $G_{jk} \neq 0$;
    that is, it equals the number of units in the network that are either connected or share a neighbor. Now, suppose that
    the degree of each unit in the network grows at the maximum rate $K_n$. In this case, all $n$ units are each correlated with $K_n^2$ other
    units. Applying Lemma~\ref{lemma:bienayme} with this number of correlated pairs yields
    \begin{equation*}
        \text{Var}\Big(r_{n_2}(\P_{\ne} - \Pe)(f_{\hat{\eta}_{S_1}} - f_{\eta_0})\Big) 
        \leq O\Big(\frac{r_n^2K_n^2}{n}\Big) \cdot \P(f_{\hat{\eta}_{S_1}} - f_{\eta_0})^2 \ .
    \end{equation*}
    \noindent 
      Using the assumption that $\sqrt{n}/K_n \leq r_n$, we have
      $O(r_n^2K_n^2/n) \cdot \P(f_{\hat{\eta}_{S_1}} - f_{\eta_0})^2  = O(\P(f_{\hat{\eta}_{S_1}} - f_{\eta_0})^2) $. Therefore, Theorem \ref{thm:as-ind-ss} must hold, so 
      $(\P_{\ne} - \Pe)(f_{\hat{\eta}_{S_1}} - f_{\eta_0}) = o_{\P}(1/r_n)$.
      which is asymptotically negligible by the given rate of the network CLT. 

\end{proof}

This particular corollary provides an example where our result holds even for
CLTs or nuisance parameters with rates not necessarily equal to $\sqrt{n}$. If
units represent areal geographies, then $G$ may represent commuting patterns
between these geographies. If each unit represents a point in space, then $G$
may denote an adjacency matrix, where units are considered ``adjacent'' if they
are within a certain distance from another unit. Similar results also exist for CLTs that may hold under other network dependency structures, such as the simpler setting where $X_i \ind X_j$ just if $G_{ij} = 0$. 

\textbf{Simulation}. We evaluate the use of cross-fitting on an
Erd\H{o}s-R\'enyi (ER) random network (with $p = 3/n$). This random network has
a maximum degree roughly bounded by $\log(n)$ \citep{Bollobas2011}, which guarantees a network CLT
will hold \citep{ogburn2022}. We estimate an ATE over the network using the
following DGP:
\begin{align*}
& G \sim \text{ER}(3/n) \ , \\
& W_1 \sim \text{Beta}(2,2), W_2 \sim \text{Pois}(10), W_3 \sim \text{Ber}(0.3) \ , \\
& A \sim \text{Ber}(\sigma(\mu(W)/20 - 1)) \ , \\
& Y \sim (2A + 1)\mu(W) + G\varepsilon_Y \ ,
\end{align*}
\noindent
where $\mu(W) = 5\mathbb{I}(W_1 > 0.4) - 2\mathbb{I}(W_1 > 0.6) +
3\mathbb{I}(W_1 > 0.7) + (2W_3 - 1)W_2 + 1$, $\sigma$ denotes the logistic
function, and $\varepsilon_Y \sim 2(\text{Beta}(6,6) - 0.5)$. Hence, the errors
of $Y$ are correlated because they result from a sum over neighbors in $G$;
$\text{Var}(Y \mid A, W) = \sigma^2(I + G^\top G)$ where $\sigma^2$ is the
variance of the Beta distribution used to construct $\varepsilon_Y$.

\begin{figure}[!h]
  \centering  \includegraphics[width=0.5\linewidth]{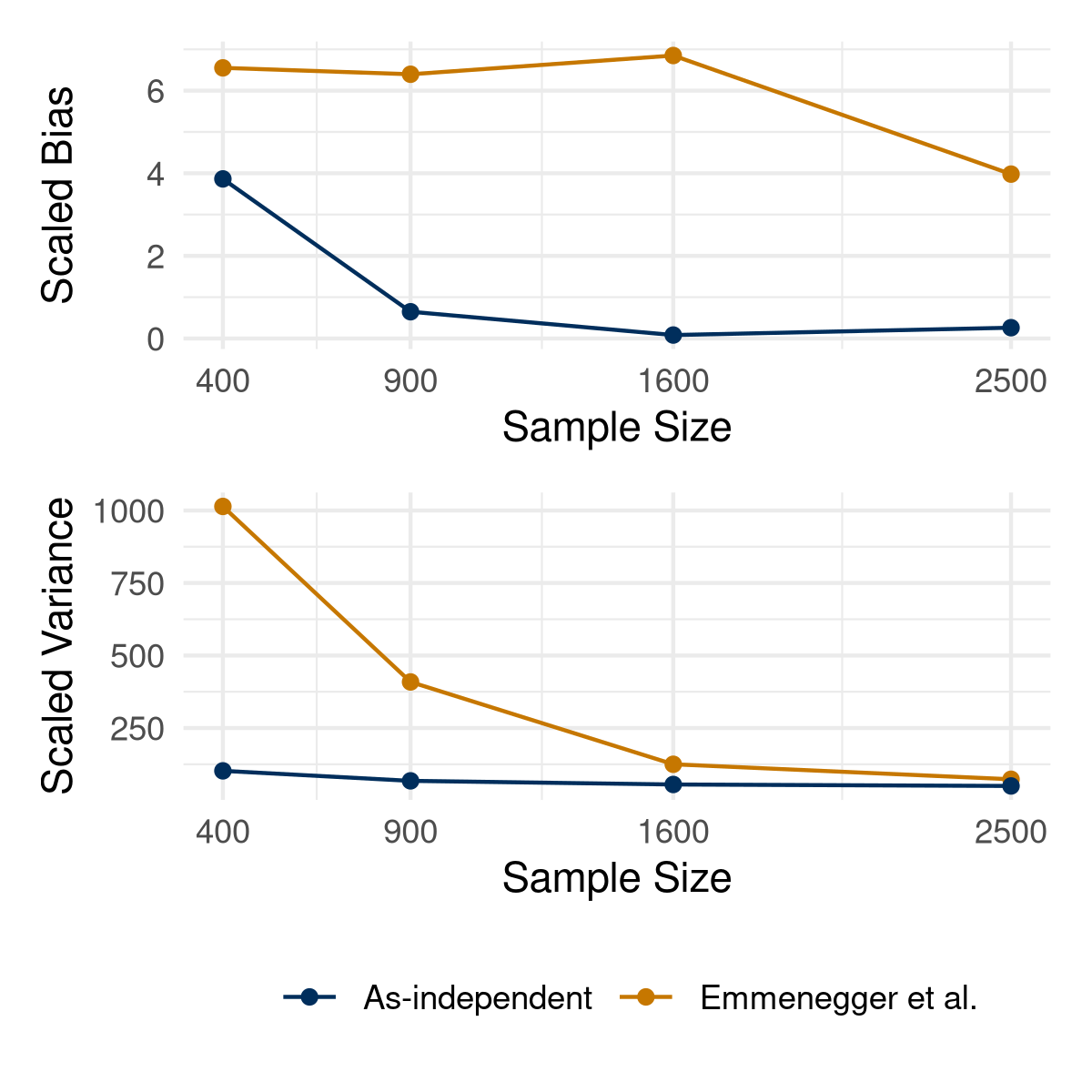}
  \caption{Operating characteristics of one-step ATE estimators in the network
  setting, comparing as-independent sample splitting to the method of \protect\cite{Emmenegger2025}.}\label{fig:network}
\end{figure}

Across 500 replicates, we estimate the ATE using a one-step estimator, with
\texttt{xgboost}~\citep{xgboost} used to learn the outcome and treatment
mechanisms---another data-adaptive algorithm generally not recognized as Donsker \citep{Smith2025}. The results of this simulation are shown in
Figure~\ref{fig:network}.

Here, as-independent sample-splitting provides a significant benefit,
especially in terms of variance. The corresponding estimator achieved low
bias and variance compared to the method of~\cite{Emmenegger2025}---any benefits from eliminating cross-fold dependence are outweighed by the loss of effective sample size from dropping correlated units. Note that this experiment needed to
be run on a relatively sparse graph because, the denser the graph, the harder
it becomes to obtain splits with no correlation. Hence, we recommend
as-independent sample-splitting in settings with complex correlation structures
such as network or spatial data.

\subsection{Time Series Data}

Finally, we consider the setting where the data consist of a single time series
$X_t$ under $m$-dependence, meaning
\begin{equation*}
    X_t \ind X_s \quad \text{whenever } \lvert t - s \rvert > m \ .
\end{equation*}

\begin{figure}[H]
\centering
\includegraphics[width=0.5\linewidth]{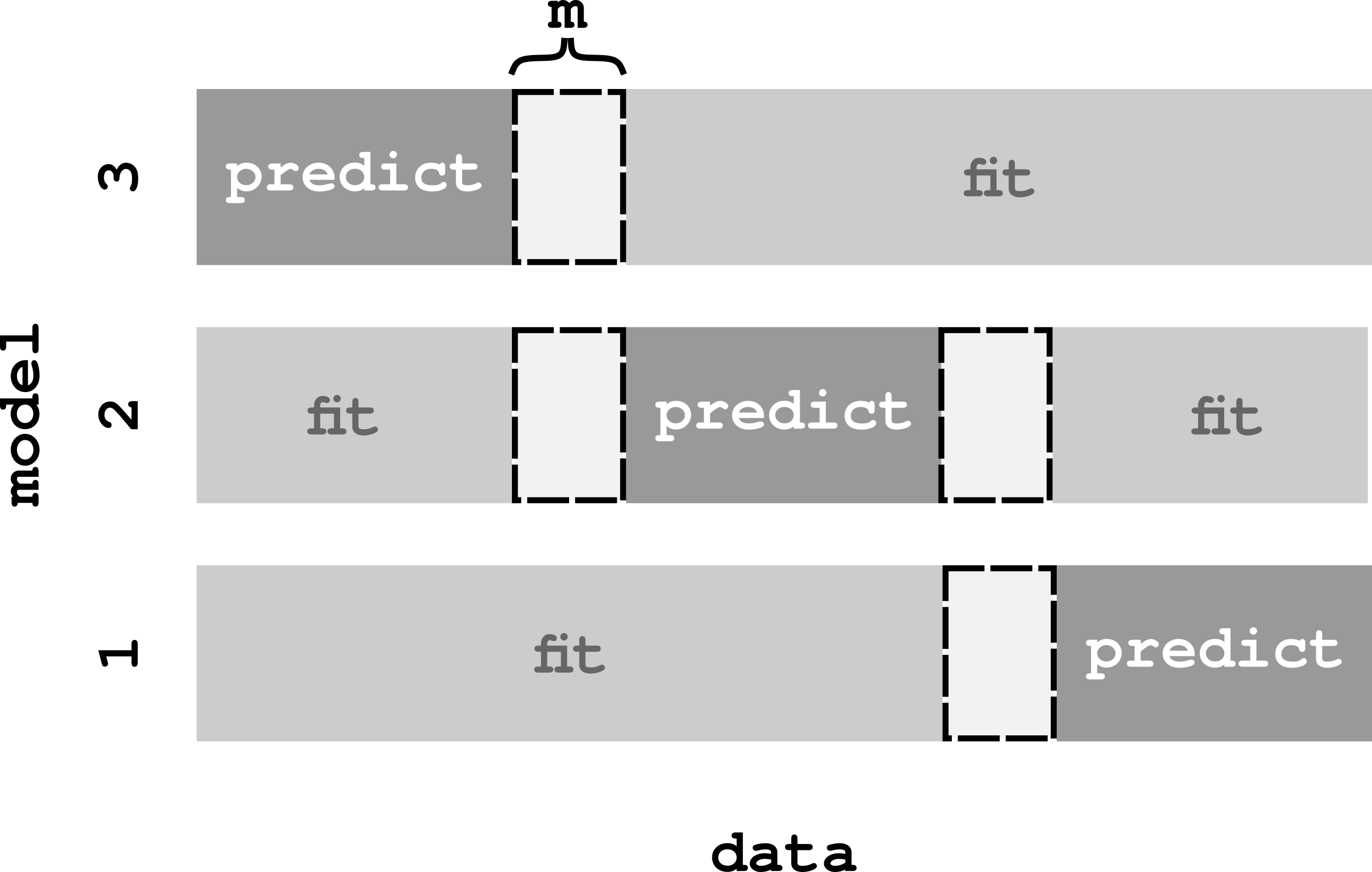}
\caption{Neighbors-left-out cross-fitting introduced by  \protect\cite{Semenova2023}.}\label{fig:nlo}
\end{figure}

Causal machine learning in this specific setting was previously discussed
by~\cite{vanderlaan2018}, though those authors do not discuss cross-fitting.
\cite{Semenova2023} also explore causal machine learning for time series,
though under a slightly different beta-mixing assumption (rather than
$m$-dependence). While those authors acknowledge that as-independent
cross-fitting can work, they propose a ``neighbors-left-out'' (NLO) method to
eliminate the negative effects of units observed close in time being strongly
correlated. NLO creates folds with gaps, in time, on either side to decrease
the dependence across separate folds (see
Figure~\ref{fig:nlo}). 

\begin{corollary}[as-IID cross-fit for fixed $m$-dependent time series]\label{cor:ts-mdep}

Let $\{X_t\}_{t=1}^T$ be a collection of units, each indexed by its timestamp, $t \in I_T = \{1, \dots, T \}$. Let Assumptions \ref{assume:clt} and \ref{assume:mse} hold under the following:
\begin{enumerate}
    \item Fixed-$m$ CLT: for some $\sigma^2$,
      $$\sqrt{T} (\P_T - \P)f_{\eta_0} \overset{D}{\to} \mathcal{N}(0, \sigma^2) \ ,$$
    (see \cite{shumway2006time}, Appendix A.2).
      \item $\{X_t\}_{t=1}^T$ is an \emph{$m$--dependent} sequence with $m$ a fixed constant, independent of the length of the series, $T$.
\end{enumerate}
\noindent Then, a cross-fit causal machine learning estimator constructed using the as-IID sample-splitting of Definition \ref{def:as-iid} satisfies, for any given sample split,
\begin{equation*}
(\P_{\ne} - \Pe) (f_{\hat{\eta}_{S_1}} - f_{\eta_0}) = o_{\P}(1/\sqrt{T}) \ ,
\end{equation*}
meaning that the empirical process bias term is 
asymptotically negligible under an \emph{$m$--dependent} time series correlation structure. 
\end{corollary}
\begin{proof}
    Let $T_2$ denote the number of units in $S_2$. To get a bound on the variance to employ Theorem \ref{thm:as-ind-ss}, we first note that by
    the fixed $m$-dependence of the sequence, each $X_t$ is correlated with at
    most $2m$ other observations. So, the total number of correlated pairs
    among the $T$ units is at most $(2m+1)T = O(T)$. Consequently, applying Lemma~\ref{lemma:bienayme},
    \begin{equation*}
        \text{Var}[\sqrt{T_2}(\P_{\ne} - \Pe)(f_{\hat{\eta}_{S_1}} - f_{\eta_0})] \leq O\Big(\frac{T}{T^2}\Big) \cdot O(T)\cdot \P(f_{\hat{\eta}_{S_1}} - f_{\eta_0})^2 = O\Big(\P(f_{\hat{\eta}_{S_1}} - f_{\eta_0})^2\Big) .
    \end{equation*}
    This satisfies the variance bound of Theorem \ref{thm:as-ind-ss}; consequently, the empirical process bias is $o_{\P}(1/\sqrt{T})$ and is therefore asymptotically negligible compared to the CLT rate. 

\end{proof}

\textbf{Simulation}. For each time point $t,$ we simulate from the following
$m$-dependent DGP, a modified version of Simulation 1 of~\cite{vanderlaan2018},
with time lag $m = 4$:
\begin{align*}
    L_{1,t}, & \quad L_{3,t} \sim \text{Ber}(0.5), \quad L_{2,t} \sim \mathrm{DUnif(1,2,3)}\\
    A_t \sim & \text{Ber}\Big(s\Big(\frac{1}{m}(X_t + \sum_{k=t-m}^{t-1}A_k - 0.5)\Big)\Big)\\
    \text{where }& X_t = \sum_{k=t-m}^{t-1}\frac{0.25}{k+m-t}\Big(2L_{1,k} +L_{2,k} - L_{3,k}\Big)\\
    Y_t \sim &  A_t + X_t + 0.3 + 20(\text{Beta}(2,2) - 0.5)
\end{align*}
Here $s(\cdot)$ is the sigmoid function, and for all $t \leq m$, $A_t \sim
\text{Ber}(0.5)$ and $Y_t \sim 4(\text{Beta}(2,2) - 0.5)$. In our simulation
experiments, we estimated a history-conditional ATE~\citep{vanderlaan2018},
learning the outcome and treatment mechanisms using generalized linear models. We provide as input to these models the $T^{1/3}$ most recent lagged values of each covariate. Since the number of covariates grows with $T$, this model does not lie in a Donsker class \citep{Belloni2012, Wainwright2019}, necessitating the use of cross-fitting.

\begin{figure}[!h]
\centering
\includegraphics[width=0.5\linewidth]{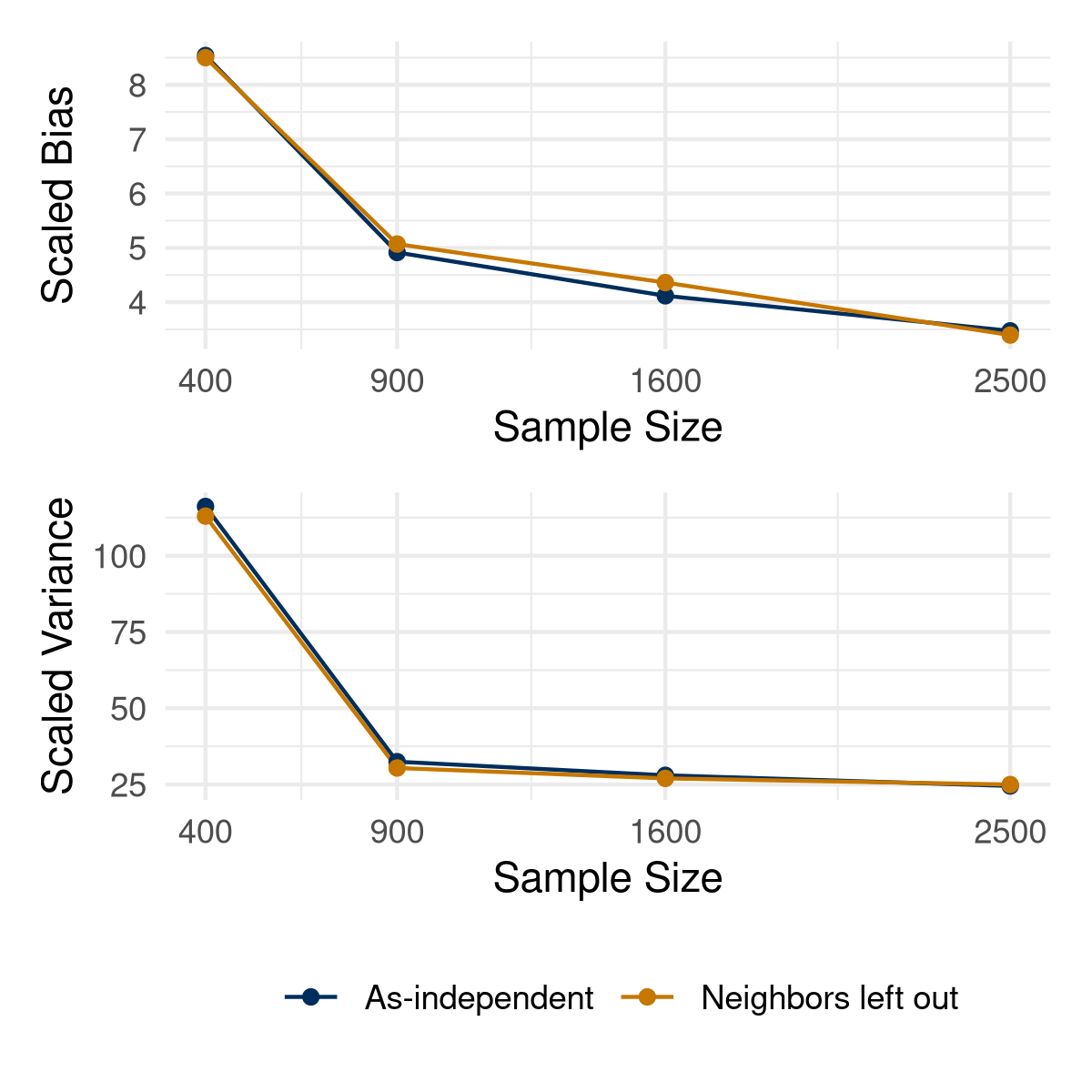}
\caption{Operating characteristics of one-step ATE estimators in the time
  series setting, comparing as-independent sample-splitting to the method of  \protect\cite{Semenova2023}.}\label{fig:time-series}
\end{figure}

We compared as-independent cross-fitting to the NLO
method of~\cite{Semenova2023}. The results are presented in
Figure~\ref{fig:time-series}. In contrast to previous methods, as-independent cross-fitting performs
almost identically in this scenario to cross-fitting designed to accommodate the correlation structure of
the data. This is likely due to the relatively lower number of correlated units in this simulation. 

\section{Concluding Remarks}\label{sec:conclusion} 

In the field of causal machine learning, recent
attention has been drawn to the use of cross-fitting to reduce reliance on
empirical process assumptions, which can prove limiting in practice by ruling
out certain families of machine learning algorithms for nuisance estimation.
Contrary to recent claims and possible intuition, we demonstrate---and
codify in our Theorem~\ref{thm:as-ind-ss}---that
cross-fitting accomplishes this goal even when units may be correlated, as in
clustered, spatial, network, or time series data.

The principal requirement for cross-fitting to work here is the existence of a central limit theorem with a variance term bounded by the variance of the estimating (or ``influence'') function $f$ whose mean $\P f$ is to be estimated. We argue that this requirement is fairly ubiquitous across typical correlated-data analyses. Hence, the limitations on the correlation structure needed for cross-fitting to work will almost certainly be satisfied by the same assumptions needed to perform estimation in general. 

In our examples, we focused on settings with a limited \textit{number} of correlated units, but our result should extend easily to the \textit{mixing} setting~\citep{Semenova2023}, in which every unit is at least somewhat correlated with every other, but dependence weakens at a given rate for units far enough apart. Similarly, we only analyzed ``single cross-fit'' estimators, in which every nuisance in $\eta$ is fitted on the same fold, but we expect the same theory to hold in the ``double cross-fit'' setting~\citep{newey2018cross} where $\eta$ consists of two nuisances, each of which is fit on a separate fold, and both are evaluated on yet another held-out fold. We leave such potential further investigations for future work.


As a pertinent reminder, we note our results only apply in the context of eliminating the empirical process term that
arises in the asymptotic analysis of causal machine learning estimators.
Correlated units can still often pose other issues in a statistical analysis. For
example, variance estimates used for the construction of confidence intervals
must be adjusted to account for inter-unit correlation. Furthermore, in some
cases the efficacy of common nuisance model selection procedures may be unclear. Work by~\cite{Davies2016} suggests that using cross-\textit{validation} for
model selection works when units are correlated. However, the performance of
individual prediction algorithms must still be considered with caution. For
example, for certain estimands in the time series setting, as-independent
splits may not be valid, as allowing prediction algorithms to, for example, use
future data to predict the past may not be sensible. 

To be sure, correlated data pose numerous challenges for both estimation
and inference. Regardless of what other challenges correlated data may pose, we aim here to
clarify a common misconception about the purpose and assumptions behind
cross-fitting in correlated data settings. We hope that having cross-fitting
available ``for free''---without the need to engineer bespoke data-splitting
strategies---will increase the ease of applying
causal machine learning techniques in settings with novel dependence structures.

\section{Code Statement}

Code to reproduce the simulation results of this work is available at the link \url{https://github.com/salbalkus/cross-fitting-dependent-data}.

\section{Acknowledgments}
The authors thank the National Science
Foundation (award no.~DGE 2140743) and the Harvard College and Harvard Data Science Initiative's
Summer Program for Undergraduates in Data Science (SPUDS) for
supporting this work. 

\bibliography{refs.bib}
\label{lastpage}

\end{document}